\RequirePackage{snapshot}
\documentclass{article}
\usepackage{color}
\usepackage{amsmath,amsfonts,amsthm}
\usepackage{multicol}
\usepackage[usenames,dvipsnames,svgnames,table]{xcolor}
\usepackage{graphicx}
\usepackage{caption}
\usepackage{subcaption}
\usepackage{float}
\usepackage{lipsum}
\usepackage{listings}
\usepackage{mathtools}
\usepackage{centernot}
\usepackage{longtable}
\usepackage{tabularx}
\usepackage{array}
\usepackage{url}
\usepackage[margin=1in]{geometry}
\usepackage[bottom]{footmisc}
\usepackage{enumerate}
\usepackage{hyperref}

\graphicspath{{graphics/}}

\begin{document}

\title{Bitcoin Transaction Graph Analysis}

\author{
 Michael Fleder \\
 \tt{mfleder@mit.edu} \\
 \and
 Michael S. Kester \\ 
 \tt{kester@eecs.harvard.edu}
 \and
 Sudeep Pillai \\
 \tt{spillai@csail.mit.edu}
}

\date{\today}

\maketitle

\section{Introduction} Bitcoins have recently become an increasingly popular
cryptocurrency through which users trade electronically and more anonymously
than via traditional electronic transfers.  Bitcoin's design keeps all
transactions in a public ledger.  The sender and receiver for each transaction
are identified only by cryptographic public-key ids. This leads to a common
misconception that it inherently provides anonymous use.  While Bitcoin's
presumed anonymity offers new avenues for commerce, several recent studies raise
user-privacy concerns.  We explore the level of anonymity in the Bitcoin system.
Our approach is two-fold: (i) We annotate the public transaction graph by
linking bitcoin public keys to “real” people - either definitively or
statistically. (ii) We run the annotated graph through our graph-analysis
framework to find and summarize activity of both known and unknown users.

\section{Contributions}
We present a bitcoin transaction-graph-annotation system in two parts.  First,
we developed a system for scraping bitcoin addresses from public forums.
Second, we include a mechanism for matching users to transactions using
incomplete transaction information.  For example, suppose we hear Bob say to
Alice: ``I sent you \$100 in bitcoins yesterday at noon''; though we don't know
the exact time of the transaction (since ``at noon'' could easily mean 11:50 or
12:10) or the exact amount in bitcoins (exchange rates fluctuate significantly),
we can generate candidate transaction matches and associated matching
probabilities.  

We also present a graph-analysis framework capable of tracing and clustering
user activity.  For example, our framework suggested the FBI seizure of Silk
Road assets as ``interesting'' activity on 10/25/2013 without prior knowledge of
the FBI or Silk Road public keys.  Furthermore, our system found close links
between Silk Road and “real” users identified with our annotation system.

\section{Background}
Recently, several research studies [3, 2, 4] have suggested the potential privacy limitations with bitcoin transactions. [3] investigates an alleged theft by leveraging external sources of information and combining them with techniques  such as context discovery and flow analysis. [4], on the other hand analyzes statistical properties of the transaction graph to answer questions about typical user behavior, spending/acquiring habits, and flow of bitcoins between multiple accounts belonging to the same user. Realizing the need for stricter privacy in the bitcoin graph, the authors in [2] suggest an extension to bitcoin that augments the protocol to allow for fully anonymous currency transactions.
\section{Threat Model}

\subsection{Attacker Goal: Tie ``real'' names to transactions } 
The ``real'' name
here may be a person's true name or username from an online public forum (or any
other public data source). The goal is to associate numerous
unrelated cryptographic IDs with an actual user.

\subsection{Attacker Capabilities} 
\begin{figure}[H] \centering 
  \begin{tabular}{cc}
    \includegraphics[width=0.45\columnwidth]{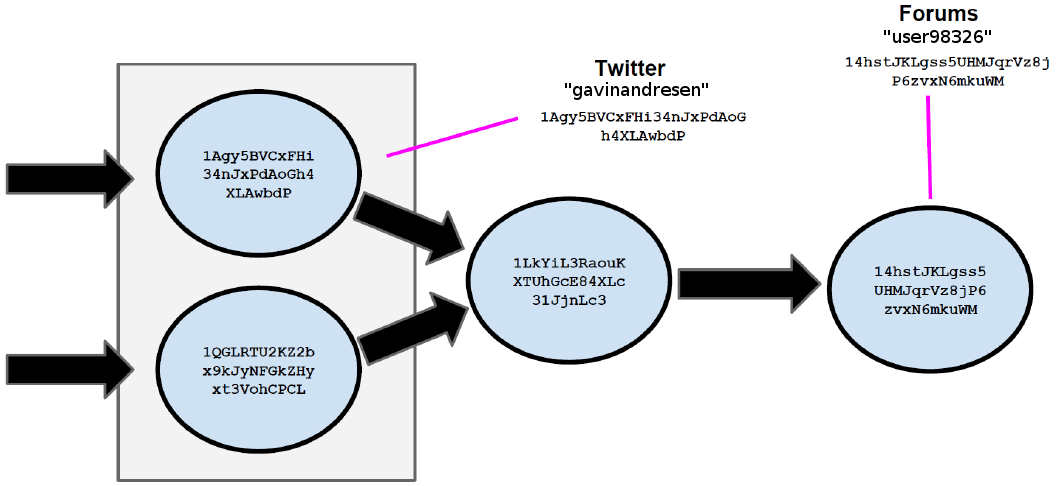}&
   \includegraphics[width=0.45\columnwidth]{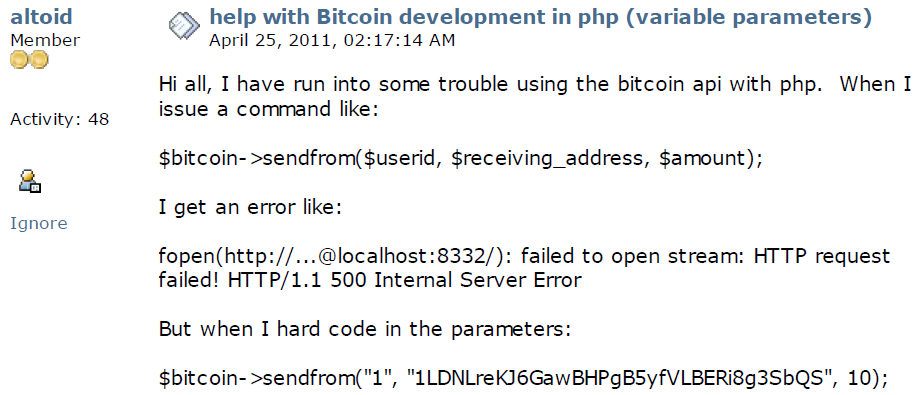}\\
   \textbf{Intentional Leak} & \textbf{Unintentional Leak}
  \end{tabular}
  \caption{On the \textbf{left}: Annotated Transaction Graph. On the \textbf{right}: Silk Road
    owner “Dread Pirate Roberts” unintentionally reveals his public key in an online forum bitcointalk.org}
  \label{fig:annotated-tx-graph}
\end{figure}

First, an attacker has access to all public
information including forums, donation sites, and public social networks from
which one can scrape bitcoin addresses that have been intentionally or
unintentionally\ref{fig:annotated-tx-graph} divulged.  That is, an attacker may
scrape (``real'' name, public key) pairs from web sites.

Second, an attacker may also ``overhear'' imprecise transaction information from
known users.  For example, an attacker may have overheard ``Alice, it’s Bob.  I sent
you \$100 bitcoins yesterday at noon.''  That is, an attacker may hear (``real''
name, some rough transaction info) pairs.

\begin{figure}[H] \centering \renewcommand{\tabcolsep}{1.5pt}
  \includegraphics[width=\columnwidth,clip,trim=20mm 50mm 20mm 50mm]{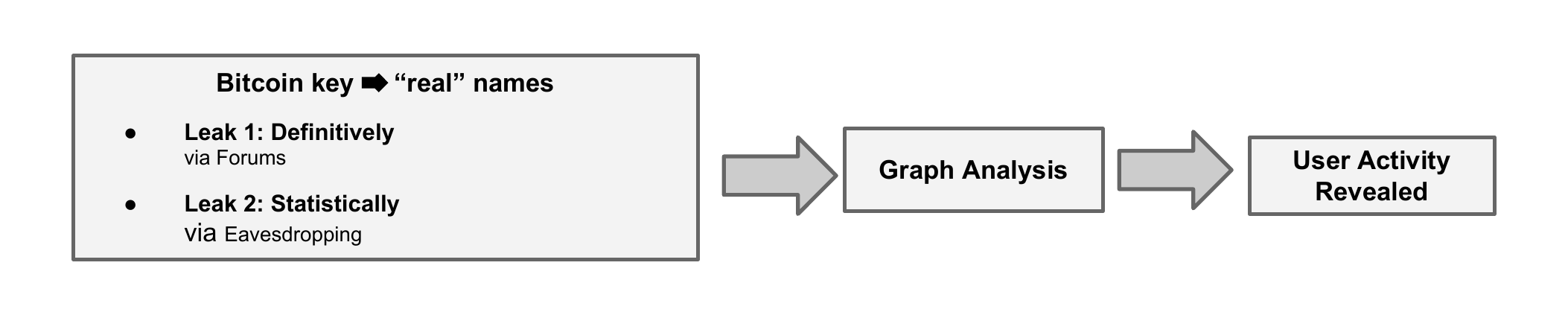}
  \caption{Attack Model}
  \label{fig:analysis-pipeline}
\end{figure}

\section{Implementation}
In this section we shall describe the several steps involved in revealing
bitcoin user activity information by leveraging publicly available transaction
information. As described in the earlier section, we investigate both,
statistical and definitive approaches. 

\subsection{Pre-processing}
As a precursor to both the above mentioned approaches, the raw transactions have
to be extracted from the full blockchain. As of Dec 13, 2013, approximately
275,000 blocks have been mined in the bitcoin block chain. Each block contains
on the order of hundreds of transactions. We describe the blockchain parsing in
the following section.
\subsubsection{Blockchain Parsing} While the standard bitcoin client
\textbf{\texttt{bitcoin-0.8.5}}\footnote{bitcoin-0.8.5
\url{http://bitcoin.org/en/download}} automatically downloads the whole
blockchain in a P2P fashion, we noticed a significantly reduced network download
rate which prompted us to download a torrent
\footnote{\url{http://sourceforge.net/projects/bitcoin/files/Bitcoin/blockchain/bootstrap.dat.torrent/download}}
from
\footnote{\url{http://sourceforge.net/projects/bitcoin/files/Bitcoin/blockchain/}}
quickly. The remaining blocks were updated automatically by the bitcoin client
after which it was indexed. While previous works \cite{reid, ron} employed a
forked version of \textbf{\texttt{bitcointools}}
\footnote{\url{https://github.com/harrigan/bitcointools}}, the newer bitcoin clients
indexed the full blockchain using LevelDB instead making the publicly available
\textbf{\texttt{bitcointools}} obsolete. Instead, we used Armory
\footnote{\url{https://github.com/etotheipi/BitcoinArmory}} to parse through the
blockchain, and wrote wrapper classes that extracted the relevant information
required to construct the transaction graph.

\subsubsection{Web Scraping}
Many users, in particular early adopters, are interested in driving bitcoin use
into more mainstream public use. One way they do this is to try to encourage
transactions. A common practice is to attach a bitcoin address as a signature
to  emails or forum posts. In forum posts especially, users contribute to the
community, for example with new mining software or a tutorial on how to get set
up to use bitcoins, and leave their address in the signature block. They expect
to receive ‘tips’ from forum readers that find their post helpful. This
practice created a natural attack vector to the anonymity of the block chain.
We can easily tie user information to transactions in the block chain.

\begin{figure}[tbh]
  \centering
    \includegraphics[width=\columnwidth,clip,trim=5mm 1mm 5mm 3mm]{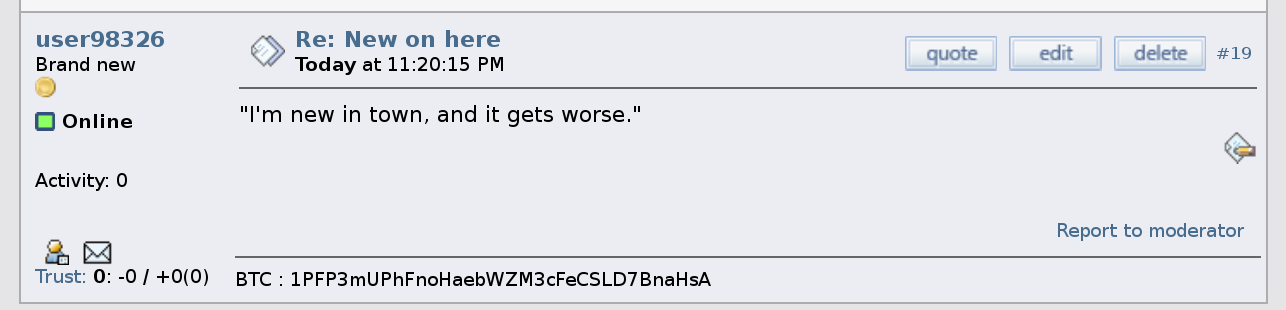}
  \caption{A typical user signature line that includes a bitcoin address for `tipping'.}
  \label{fig:voodah_post}
\end{figure}

We used a python package called Scrapy\footnote{http://scrapy.org/} to fetch
and parse the forum pages(fig.~\ref{fig:voodah_post}). We wrote a spider that crawls
bitcointalk.org in a breadth-first manner looking for post
signatures that might contain bitcoin addresses (i.e. it
matched the regular expression r`1.\{26,33\}'). We then took this string and
verified that it was a legitimate bitcoin public key (bitcoin addresses include
a built-in checksum) to avoid attempting to annotate a large number nodes that
can't possibly appear in the blockchain. 

We were able to find a large number of forum users that can be directly linked
to their keys in the transaction graph. We ran the scraping code for just under
30 hours. During this time we followed links up to four deep from the home
page. This covered a total of 44,086 pages and 89,088 posts that included a
valid bitcoin address. Of this, we found 2,322 unique users with 2,404 addresses
that passed our validation.

\subsection{Transaction Fingerprinting}
Here we analyze the difficulty in taking rough information regarding transaction
time and value, and matching it to an exact transaction in the blockchain.  For
example, if we overhear Bob telling Alice, “I sent you \$100 USD yesterday at
noon,” we examine the difficulty of finding a matching transaction in the
blockchain.  Suppose we assume the value of bitcoins fluctuated \$1 USD
yesterday, and that the noon timestamp is accurate to within 5 minutes.  Then we
have a number of candidate transactions to examine.  Continuing with this
example, we convert from USD to bitcoins using daily market prices from
BlockChain\footnote{https://blockchain.info/charts/market-price, December 12,
2013}.  Next, we examine all transactions occurring in both ranges of [\$99,
\$101] and [11:55 AM, 12:05 PM].

To generalize this example, we examine every transaction in the block chain, and
then create time and USD value windows by varying amounts to see how many other
transactions will match this weaker, window criteria.  The figure below shows,
for given USD and time windows, the average number of transactions that will
match any particular transaction.  Bitcoins have over time become more popular
and frequently traded over time, so more candidate transactions will match the
given dollar and time criteria in recent months.

\begin{figure}[H] \centering \renewcommand{\tabcolsep}{1.5pt}
  \begin{tabular}{c}
    \includegraphics[width=0.6\columnwidth]{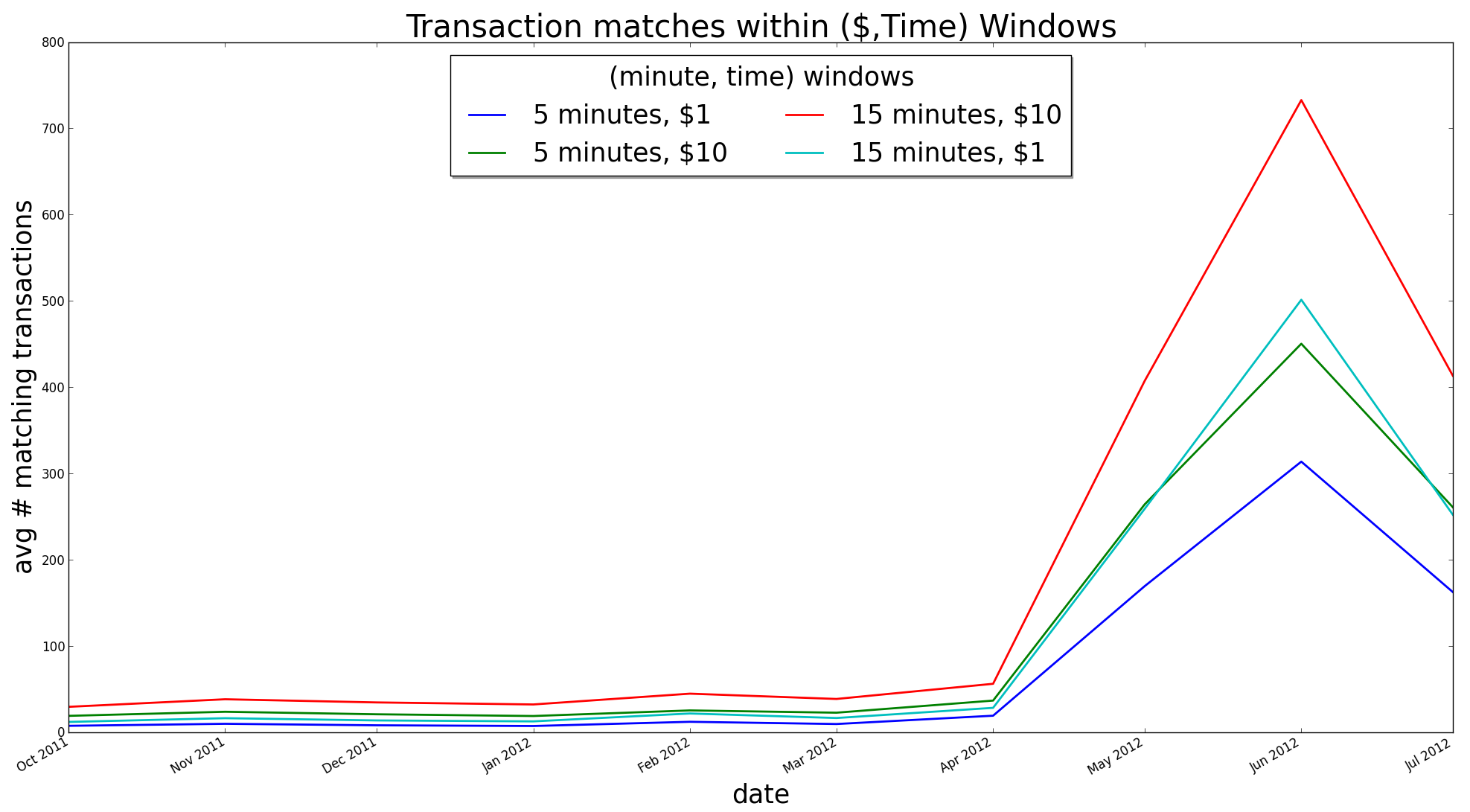} 
  \end{tabular}
  \caption {Transaction ambiguity resulting from inexact time and
  approximate USD worth}
  \label{fig:fingerprinting}
\end{figure}

Continuing with the example from the start of this section, we may identify
Bob’s public key with probability ~ 1/10 assuming the conversation took place in
March 2012.

Whether by using this fingerprinting tool or by scraping data from online
sources, we are able to annotate the blockchain with additional,
user-identifying information.  In the former case, the annotations may have
associated probabilities.

\subsection{Graph Analysis}
\begin{figure}[H]
  \centering
  \renewcommand{\tabcolsep}{1.5pt}
  \begin{tabular}{c}
    \includegraphics[width=0.75\columnwidth,clip,trim=0mm 5mm 0mm 5mm]{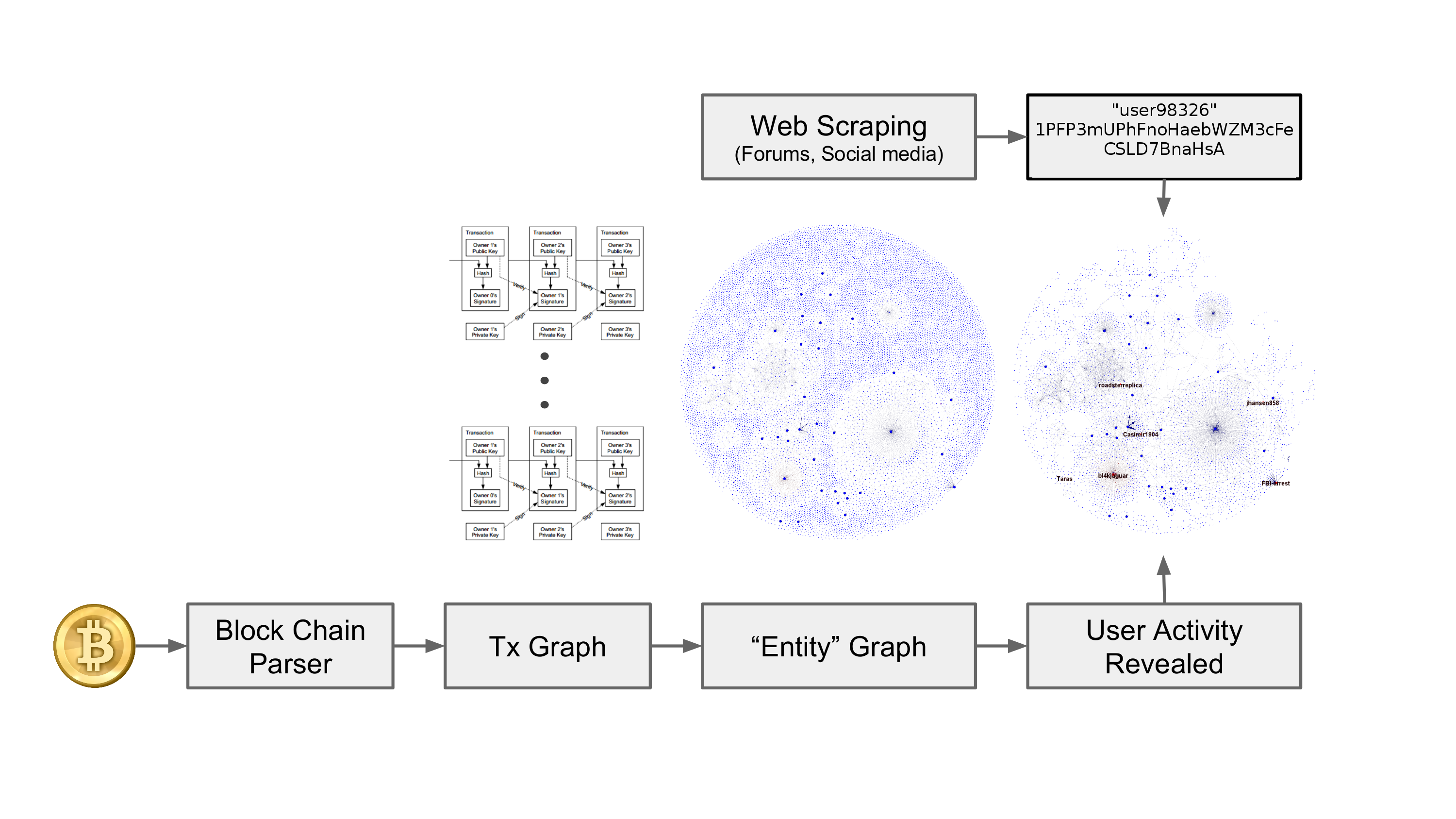}
  \end{tabular}
  \caption{The graph analysis pipeline that was used to reveal user identity
    constructs a user network graph as shown, and annotates the users in the
    graph with web scraped results. }
  \label{fig:graph-analysis-pipline}
\end{figure}

We developed a graph analysis framework in order to de-anonymize users' identities
given publicly available information such as scraped bitcoin forum users, and
bitcoin transaction information. Figure~\ref{fig:graph-analysis-pipline}
outlines the different components of our framework.

\clearpage
\subsubsection{The Transaction Graph}
Once the transaction records are extracted from the blockchain, we construct a
transaction graph that gives an intuition towards the flow of bitcoins between
public key addresses over time. More specifically, the transaction graph is a
directed graph where the nodes denote public addresses of anonymous individuals
or ``entities'' and the directed edge represents a particular transaction from a
source address to a target address. Since both the source and the target
``entities'' can arbitrarily generate new public-private key pairs for each
subsequent transaction, many public key addresses may only appear once or a few
times in the transaction graph. Additionally, typical transactions in today's
blockchain are multi-input/multi-output transactions. For a more detailed
reference, please refer to \cite{nakamoto2008bitcoin}. We use similar techniques
as described in \cite{reid}, to prune transactions in our graph. For our experiments, we construct the
transaction graph for a 24-hour period on October, 25th, 2013. The transaction
graph consists of 89,806 transactions, with 80,030 unique vertices (or public
key addresses). We also constructed a transaction graph for a 7 month period
consisting of 1,669,728 transactions,
spanning the months of Mar 2013 through Oct 2013, in an attempt to reveal any
links between the bitcoin forum users and the Silk Road nodes, before it was
shut down. 

\subsubsection{The User Graph}
In this section, we focus on the 1-day transaction graph constructed and
describe our findings on particular activities that are immediately visible
after our graph analysis algorithm is applied. Using the transaction graph, we
construct a proxy directed graph called the user graph U similar to that
described in \cite{reid}, where the user or entity consists of a collection of public
key addresses that were used during separate transactions. As noted in
\cite{nakamoto2008bitcoin}, we link together transactions with multi-inputs as originating
from the same user. This allows the creation of a user graph by performing a
transitive closure on the set of public key address involved in all
transactions, after the multi-input public keys are linked together. We use
existing tools \footnote{\url{http://compbio.cs.uic.edu/data/bitcoin/}} to construct
the entity/user graph, where the vertices now represent physical users/entities,
and edges rerepsent a transaction between a source user and a target user. As we
construct the user graph from a transaction graph spanning a 24-hour period, our
user netowrk is not quite indicative of the true user network as several public
key addresses that may have appeared before or after the 24-hour transaction
period are not used to link addresses. The resulting user network for the Oct
25, 2013 consists of 54,941 users with 89,806 edges. 

\begin{figure}[H]
  \centering
  \renewcommand{\tabcolsep}{1.5pt}
  \begin{tabular}{c}
    \includegraphics[width=0.8\columnwidth,clip,trim=10mm 10mm 15mm 10mm]{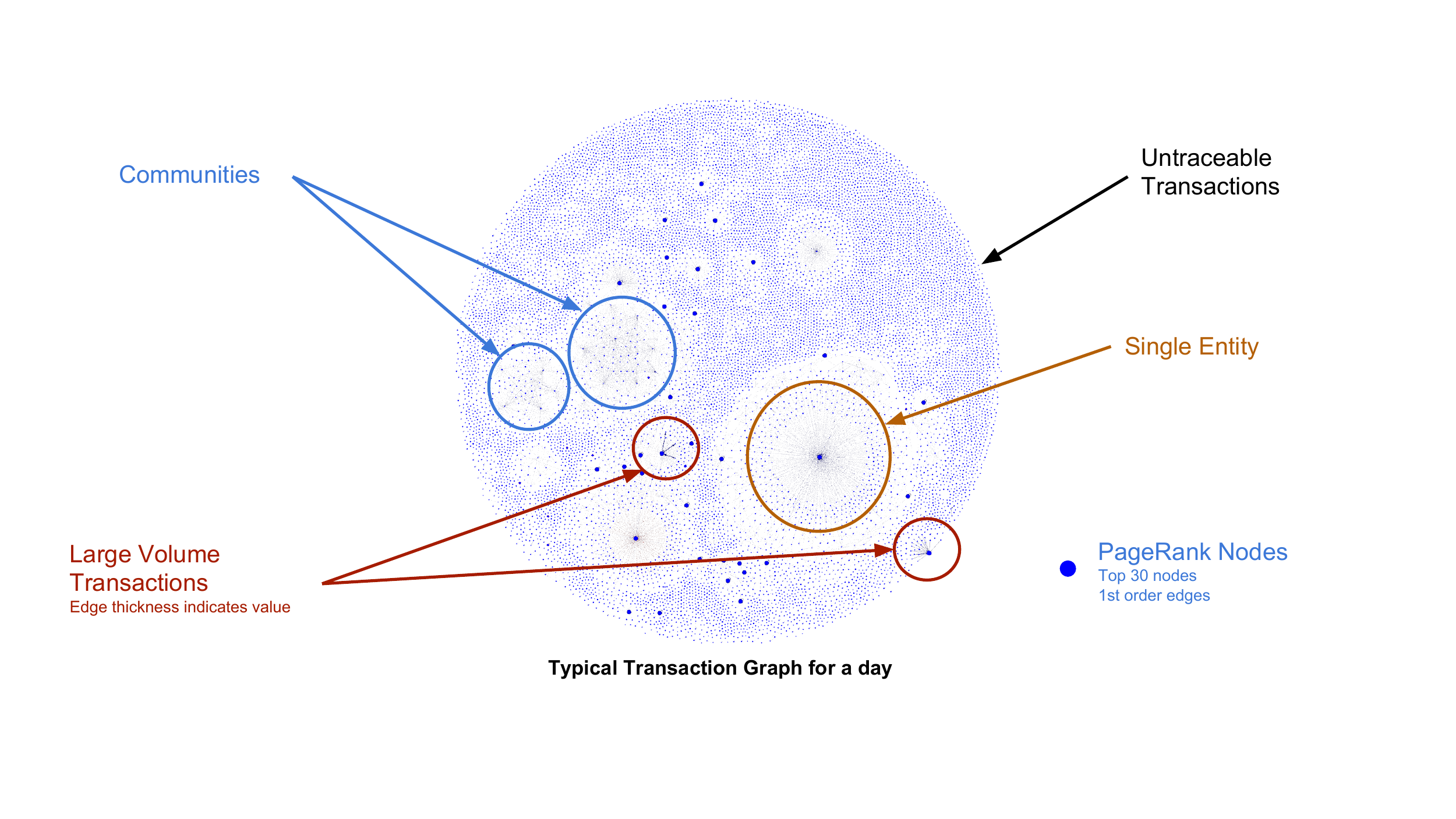}
  \end{tabular}
  \caption{An ``entity'' graph for a typical day (Oct 25, 2013) labeled with the top 30
page ranked nodes. While a significant set of users are untraceable, several
different activities and network layouts are noticeable such as communities,
single entities, and large volume transactions. }
  \label{fig:graph-page-rank}
\end{figure}

\subsection{Page Rank}
Due to the nature of bitcoin transactions in the directed user graph, we see a
direct resemblance of this graph to those constructed by search-engines. Most
search-engines, in particular Google, use PageRank as a metric to rank websites
based on their importance. Intuitively, the algorithm prefers nodes in a
directed graph that are most easily reached, or in our case, nodes that receive
a large enough traffic to be labeled as important. We use PageRank as a guide to
determine the most intersting nodes, or users in our user graph to further
investigate their linkage with known forum
users. Figure~\ref{fig:graph-page-rank} shows the user graph for
Oct 25, 2013, and labels the top page ranked nodes with a larger node
size. As expected, most of the users in the graph are not connected implying
that these nodes are not of much importance as they are not traceable to other
users.  One also notices several other types of activities or transactions
involving communities, single entities, and even large volume transactions as
indicated by the thickness of the edges. Given some of the most popular public
key addresses from BlockChain.info\footnote{\url{https://blockchain.info/}}, we were able to determine that one of the
single entity nodes with a high in and out-degree was in fact a Bitcoin gambling website
called SatoshiDICE\footnote{\url{http://legacy.satoshidice.com/}}. 

\subsection{User De-anonymization} An activity of particular interest was the
seizure of Silk Road funds to a publicly known address of the FBI\footnote{FBI:
1FfmbHfnpaZjKFvyi1okTjJJusN455paPH} via 445 transactions of exactly 324
Bitcoins. Our graph analysis algorithm picked out this specific FBI address as a
user of high importance (high page-ranked node). This validates our algorithm to
appropriately pick out nodes of particular interest, and allows for further
investigation of these high page-ranked nodes. Given this information and
web-scraped bitcoin forum user's information, we were also able to backtrack
from these transactions to uncover bitcoin forum users that were only a single
hop away from the Silk Road nodes. This implies that the bitcoin forum users
transacted with some user that had directly transacted with the Silk Road. Due
to DPR's\cite{rondid} arrest earlier that month, we analysed transactions up to 7
months before the seizure (Mar 25, 2013 through Oct 25, 2013). We were also able
to uncover direct transactions from multiple bitcoin forum users to
SatoshiDICE\footnote{SatoshiDICE 48\%: 1dice8EMZmqKvrGE4Qc9bUFf9PX3xaYDp},
implying that they may have gambled at some point during that 7 month
period. Interestingly enough we were also able to find direct transactions to
Wikileaks\footnote{\url{http://wikileaks.org/} : 1HB5XMLmzFVj8ALj6mfBsbifRoD4miY36v}
from a few of the bitcoin forum users.

\begin{figure}[H]
  \centering
  \renewcommand{\tabcolsep}{1.5pt}
  \begin{tabular}{cc}
    \includegraphics[width=0.75\columnwidth,clip,trim=0mm 5mm 0mm 5mm]{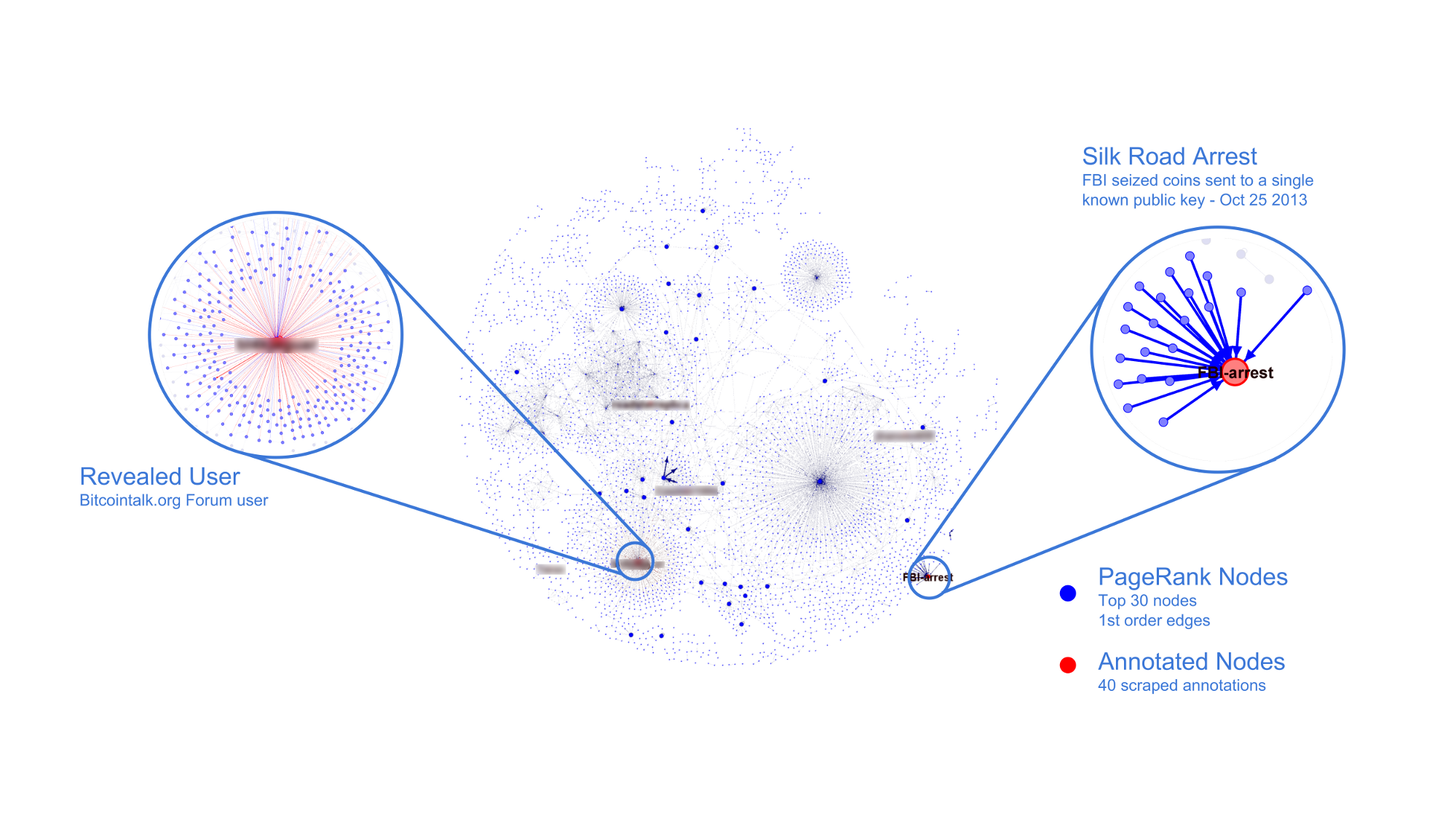}
  \end{tabular}
  \caption{The transaction graph for Oct 25, 2013 showing the top page ranked nodes and
    their first order edges with annotations from web-scraped results. Several
    noticeable activities, including the seizure of bitcoins from Silk Road entities to a
    single known FBI address, tend to be involved with the top page ranked nodes. }
  \label{fig:graph-annotations-fbi}
\end{figure}

\section{Conclusion} 
In conclusion, we showed that by leveraging several sources
of publicly available information via web-scraped forums and Bitcoin's
transaction ledger, the bitcoin transaction network is shown to be not entirely anonymous. Furthermore,
we were able to tie bitcoin forum users with the original Silk Road nodes with
only a single intermediary. We were also able to successfully find transactions
that directly linked the scraped bitcoin forum users with known entities like
SatoshiDICE, and Wikileaks implying that they may have dealt with, supported, or
interacting with such entities.


\clearpage

\bibliographystyle{plain}
\bibliography{references}

\end{document}